\documentclass[aps,pra,preprint,groupedaddress]{revtex4}
\usepackage{amssymb,bm}
\usepackage[dvips]{graphicx}

\begin{document}
\bibliographystyle{apsrev}

\title{Spin effects in $p\bar p$ interaction and their possible use
to polarize   antiproton beams.}
\author{V.F. Dmitriev }
\email{V.F.Dmitriev@inp.nsk.su}
\author{A.I. Milstein  }
\email{A.I.Milstein@inp.nsk.su}
\author{V.M. Strakhovenko}
\email{V.M.Strakhovenko@inp.nsk.su} \affiliation{Budker Institute
of Nuclear Physics, 630090 Novosibirsk, Russia}

\date{\today}

\begin{abstract}
Low energy $p\bar p$ interaction is considered taking into account
 the polarization of both particles. The corresponding cross
sections are calculated using  the Paris nucleon-antinucleon optical
potential.  Then they are   applied to the analysis  of the
polarization buildup which is due to the interaction of stored
antiprotons with  polarized protons of a hydrogen target. It is
shown  that,  at realistic  parameters of a storage ring and a
target,  the filtering mechanism provides a noticeable polarization
in a time comparable with the beam lifetime.

\end{abstract}

\maketitle

\section{Introduction}
An extensive physical  program with polarized antiprotons has been
proposed recently by the PAX Collaboration \cite{PAX05}. This
program  has initiated a wide discussion on the possibility to  use
a polarized hydrogen gas target to polarize stored antiprotons (see
\cite{PAX05, Rathman05} and literature therein). Various
modifications of the filtering method first proposed in Ref.
\cite{Csonka68} have been considered. The filtering method exploits
the dependence of the scattering cross section on the orientations
of the target and projectile proton (antiproton) spins. Due to this
dependence beam protons  with positive spin projections on the
direction of the target polarization  scatter out of the beam (
scattering angle $\vartheta$ larger than the acceptance angle
$\theta_{acc}\ll 1$) at a rate different from protons with a
negative spin projection. As a result the beam becomes polarized.
This method can also be used for the antiproton beam.

In Refs.\cite{MilStr05,Kolya06}, it was shown that , for
$\vartheta<\theta_{acc}$ (a proton or antiproton remains in the
beam), the polarization buildup is completely due to the spin-flip
transitions. The corresponding cross sections turn out to be
negligibly small for both proton-proton and proton-electron
scattering. For a  pure electron target, filtering mechanism  also
does not provide  a noticeable polarization, Ref.\cite{MilStr05}.
Thus, it is necessary to study in detail the filtering method for
the antiproton beam using a hydrogen gas target with the proton
polarization.

The method suggested in  \cite{Csonka68} has been realized in the
experiment \cite{Haeberli93}, where  23-MeV stored protons scattered
on an internal gas target of the polarized hydrogen atoms. The
measured polarization degree is in reasonable agreement with the
theoretical predictions based on the known $pp$ scattering cross
section , see discussion in Refs.\cite{MilStr05,Kolya06,Rathman05}.
Theoretical prediction for the rate of polarization buildup of the
antiproton beam is essentially more complicated problem because the
spin-dependent part of the $p\bar p$ scattering cross section is not
well known both theoretically and experimentally. Theoretical
evaluation  of this cross section  contains some uncertainties
because nowadays QCD can not describe quantitatively the low energy
nucleon-antinucleon interaction. Therefore, it was necessary to
apply  various phenomenological approaches  in order to explain
numerous experimental data , see , e.g., Refs.
\cite{pot82,bonn91,bonn95,paris82,paris94,Partial94} and recent
reviews \cite{KBMR02,KBR05}. However, some parameters of the models
determining the spin-dependent part of the cross sections are not
well defined from the experimental data available \cite{Rich94}.

In the present paper, we calculate the  cross section of polarized
proton and antiproton interaction using the Paris
nucleon-antinucleon optical potential $V_{N\bar N}$ with the
parameters given in Refs.\cite{paris82,paris94,paris99}. This
potential
 has the form:
\begin{equation}\label{VNN}
V_{N\bar N} = U_{N\bar N}-i\, W_{N\bar N} \, ,
\end{equation}
where the real part $U_{N\bar N}$  is the  $G$-parity transform of
the well established Paris $NN$ potential for the long- and
medium-ranged distances ($r\gtrsim 1\,\mbox{fm}$), and some
phenomenological part for  shorter distances. The absorptive part,
$W_{N\bar N}$, of the optical potential takes into account the
inelastic channels of $N\bar N$ interaction, i.e. annihilation into
mesons, being important at short distances.  Our knowledge of
$W_{N\bar N}$ is more restricted than that of $U_{N\bar N}$.

\section{Cross sections}
It is convenient to calculate the cross sections of $p\bar p$
scattering in the center-of-mass frame where antiproton and proton
have the momenta $\bm p$ and $-\bm p$, respectively. We assume that
$p\ll M$ ( $M$ is the proton mass) and perform calculations in the
nonrelativistic approximation,  so that the momentum of antiproton
in the lab frame is $\bm p_{lab}=2\bm p$.  Let us direct the polar
axis $z$ along the unit vector ${\bm \nu}={\bm p}/p$. Kinetics of
polarization depends on the cross section
\begin{equation}
\sigma=\sigma_{ann}+\sigma_{cex}+\sigma_{el}\,,
\end{equation}
where $\sigma_{ann}$   is the annihilation cross section,
$\sigma_{cex}$ is the charge-exchange cross section of the process
$p\bar p\to n\bar n$ , and $\sigma_{el}$ is the elastic cross
section of $p\bar p$ scattering summed up over final spin states,
integrated over the  azimuth angle, and over the scattering angle
$\vartheta$ from the acceptance angle $\theta_{acc}$ to $\pi$. We
remind that $\theta_{acc}$ is defined in the center-of-mass frame.
In the lab frame the acceptance angle is
$\theta_{acc}^{(l)}=\theta_{acc}/2$ . The cross section $\sigma$ has
the form
\begin{eqnarray}\label{sigma}
\sigma=\sigma_0+({\bm\zeta}_1\cdot{\bm\zeta}_2)\,\sigma_1+
({\bm\zeta}_1\cdot{\bm\nu})({\bm\zeta}_2\cdot{\bm\nu})\,(\sigma_2
-\sigma_1) \, ,
\end{eqnarray}
where ${\bm\zeta}_1$ and ${\bm\zeta}_2$ are the unit polarization
vectors of the proton and antiproton, respectively. For example,
$\sigma=\sigma_0+\sigma_2$ for ${\bm\zeta}_1\|{\bm\zeta}_2$ and
${\bm\zeta}_1\|\bm \nu$ ;
  $\sigma=\sigma_0+\sigma_1$ for
${\bm\zeta}_1\|{\bm\zeta}_2$ and  ${\bm\zeta}_1 \perp\bm \nu$.

 We choose the
quantization axes along the vector $\bm\nu$ and express the spin
wave function of the initial $p\bar p$ state via the spin wave
function, corresponding to the total spin $S=1$ and projection
$S_z=\mu$,  and the spin wave function, corresponding to the total
spin $S=0$. Then we obtain
\begin{eqnarray}\label{Sigma}
\sigma_0&=&\frac{1}{2}\Sigma_{11}+\frac{1}{4}(\Sigma_{10}+
\Sigma_{00})\,,\nonumber\\
\sigma_1&=&\frac{1}{4}(\Sigma_{10}-\Sigma_{00})\,,\nonumber\\
\sigma_2&=&\frac{1}{2}\Sigma_{11}-\frac{1}{4}(\Sigma_{10}+\Sigma_{00})\,,
\end{eqnarray}
where $\Sigma_{1\mu}$ is the cross section calculated for the total
spin $S=1$ and $J_z=\mu$,  $J_z$ is the projection of the total
angular momentum,  $\Sigma_{00}$ is the cross section calculated for
the total spin $S=0$ (so that $J_z=0$). Note that
$\Sigma_{11}=\Sigma_{1-1}$.

Our potential is a sum of the  nucleon-antinucleon optical potential
$V_{N\bar N}$ and the Coulomb potential $V_C(r)=-e^2/r$, where $e$
is the proton charge. The hadronic amplitude  can be represented as
a sum of a pure electromagnetic amplitude and the  strong amplitude.
We emphasize that the latter  does not coincide with the strong
amplitude calculated without account for the electromagnetic
interaction. The strong amplitude is not  singular at small
scattering angle $\vartheta$. In the nonrelativistic  limit, the
triplet $F_{C1}$ and singlet $F_{C0}$ electromagnetic amplitudes
coincide with the amplitude $f_{C}(\vartheta)$ where
\begin{eqnarray}{\label{eq:MChat}}
&& f_{C}(\vartheta)=\frac{\alpha
}{4vp\sin^2(\vartheta/2)}\exp\{i(\alpha/v)\ln[\sin(\vartheta/2)]+2i\chi_0\}\,
,\nonumber\\
&&\chi_0=\arg\Gamma(1-\frac{i\,\alpha}{2v})\, ,
\end{eqnarray}
where $v=p/M$ is the nucleon (antinucleon) velocity in the  center-of-mass frame,
 $\alpha=e^2$ is the fine structure constant.
It was shown in Ref.\cite{MilStr05} that, for time comparable with
the  beam life-time, the polarization degree of the antiproton beam,
$P_B$, is determined by the ratio $\sigma_{1,2}/\sigma_0$.
 The pure Coulomb cross section is spin independent
and, therefore, contributes only to $\sigma_0$ that diminishes the
ratio $\sigma_{1,2}/\sigma_0$. Therefore, we  do not consider very
small c. m. energies $E<5$ MeV because in this region the Coulomb
cross section becomes essentially larger than the nuclear cross
section. Note that the interference of the Coulomb amplitude and the
spin-dependent part of the strong amplitude, corresponding to
elastic scattering without spin flip, may be important
\cite{Meyer94}.

In our calculation, we use the standard partial-wave analysis (see,
e.g., Refs.\cite{paris94,Partial94}). For the strong  elastic
triplet scattering amplitude ($p\bar p\to p\bar p$), we have
\begin{eqnarray}{\label{Tripl}}
&&F_{1\mu}^{el}=\frac{i\sqrt{4\pi}}{2p}\sum_{m,\,L,\,J}
Y_{L\,m}(\vartheta,\varphi)C^{J\mu}_{Lm,1\mu-m}R^{J}_{L\mu}\, ,\nonumber\\
&&R^{J}_{L\mu}=\sum_{L'}(-1)^{\frac{L-L'}{2}}
\sqrt{2L'+1}C^{J\mu}_{L'0,1\mu}\exp(i\chi_L+i\chi_{L'})
\left(\delta_{LL'}-S^{J}_{LL'}\right)\, ,\nonumber\\
&&\chi_L=\arg\Gamma(L+1-\frac{i\,\alpha}{2v})\, ,
\end{eqnarray}
where summation over $L,\,L'$ is performed under the conditions
$L,\,L'=J,\,J\pm 1$ and $|L-L'|=0,\,2$. The strong singlet elastic
scattering amplitude is given by
\begin{eqnarray}{\label{Singl}}
F_0^{el}=\frac{i\sqrt{4\pi}}{2p}\sum_{L}
\sqrt{2L+1}\,Y_{L\,0}(\vartheta,\varphi)\exp(2i\chi_L)
\left(1-S_{L}\right)\,.
\end{eqnarray}
In Eqs.(\ref{Tripl}) and (\ref{Singl}), $S^{J}_{LL'}$ and $S_{L}$
are the partial elastic triplet and singlet scattering amplitudes,
respectively, $Y_{L\,m}(\vartheta,\varphi)$ are the spherical
functions, $C^{J\mu}_{Lm,1\mu-m}$ are Clebsch-Gordan coefficients.

The  charge exchange amplitude ($p\bar p\to n\bar n$) is given by
\begin{eqnarray}{\label{Triplcex}}
&&F_{1\mu}^{cex}=-\frac{i\sqrt{4\pi}}{2p}\sum_{m,\,L,\,J}
Y_{L\,m}(\vartheta,\varphi)C^{J\mu}_{Lm,1\mu-m}{\tilde R}^{J}_{L\mu}\, ,\nonumber\\
&&{\tilde R}^{J}_{L\mu}=\sum_{L'}(-1)^{\frac{L-L'}{2}}
\sqrt{2L'+1}C^{J\mu}_{L'0,1\mu}\exp(i\chi_{L'})\,{\tilde
S}^{J}_{LL'}\, ,
\end{eqnarray}
for the triplet contribution and
\begin{eqnarray}{\label{Singlcex}}
F_0^{cex}=-\frac{i\sqrt{4\pi}}{2p}\sum_{L}
\sqrt{2L+1}\,Y_{L\,0}(\vartheta,\varphi)\exp(i\chi_L) \,{\tilde
S}_{L}
\end{eqnarray}
for the singlet one. In Eqs.(\ref{Triplcex}) and (\ref{Singlcex}),
${\tilde S}^{J}_{LL'}$ and ${\tilde S}_{L}$ are the partial charge
exchange triplet and singlet scattering amplitudes, respectively.

The cross sections $\Sigma_{1\mu}$ and $\Sigma_{00}$ can be
represented as a sum of pure Coulomb contributions,
$\Sigma_{1\mu}^C$ and $\Sigma_{00}^C$, hadronic contributions,
$\Sigma_{1\mu}^h$ and $\Sigma_{00}^h$, and the interference
terms, $\Sigma_{1\mu}^{int}$ and $\Sigma_{00}^{int}$. For Coulomb
contributions, we have
\begin{eqnarray}{\label{CC}}
\Sigma_{1\mu}^C=\Sigma_{00}^C=\sigma^C=\frac{\pi\alpha^2}{(vp\theta_{acc})^2}\,,
\end{eqnarray}
where  a smallness of $\theta_{acc}$ is taken into account. It
follows from the optical theorem that the total hadronic cross
sections are
\begin{eqnarray}{\label{Hard}}
&&\Sigma_{1\mu}^h=\frac{2\pi}{p^2}\sum_{L,\,J}\sqrt{2L+1}
\,C^{J\mu}_{L0,1\mu} \mbox{Re}\,R^{J}_{L\mu}\,
,\nonumber\\
&&\Sigma_{00}^h=\frac{2\pi}{p^2}\sum_{L} (2L+1)\,
\mbox{Re}\left[\exp(2i\chi_L)(1-\,S_{L})\right]\,.
\end{eqnarray}
The terms in the  cross section corresponding to the interference of
the Coulomb and the strong elastic $p\bar p$ amplitudes read
\begin{eqnarray}{\label{int}}
&&\Sigma_{1\mu}^{int}=-\frac{2\pi\alpha}{vp^2}\log\left(\frac{2}{\theta_{acc}}\right)
\sum_{L,\,J}\sqrt{2L+1} \,C^{J\mu}_{L0,1\mu}\nonumber\\
&&\times\left\{
\mbox{Im}\bigg[\exp(-2i\chi_0)\,R^{J}_{L\mu}\bigg]+\frac{\alpha}{2v}
\log\left(\frac{2}{\theta_{acc}}\right)
\mbox{Re}\bigg[\exp(-2i\chi_0)\,R^{J}_{L\mu}\bigg]\right\}\,
,\nonumber\\
&&\Sigma_{00}^{int}=-\frac{2\pi\alpha}{vp^2}\log\left(\frac{2}{\theta_{acc}}\right)\sum_{L}
(2L+1)\, \Big\{\mbox{Im}\bigg[\exp\bigg(2i(\chi_L-\chi_0)\bigg)(1-\,S_{L})\bigg]\nonumber\\
&&+\frac{\alpha}{2v}
\log\left(\frac{2}{\theta_{acc}}\right)\mbox{Re}\bigg[\exp\bigg(2i(\chi_L-\chi_0)\bigg)
(1-\,S_{L})\bigg]\Big\}\,.
\end{eqnarray}
The hadronic contributions to the elastic cross section of $p\bar
p\to p\bar p$ process  have the form:
\begin{eqnarray}{\label{elastic}}
&&\Sigma_{1\mu}^{el}=\frac{\pi}{p^2}\sum_{L,\,J}\left|R^{J}_{L\mu}\right|^2\,
,\nonumber\\
&&\Sigma_{00}^{el}=\frac{\pi}{p^2}\sum_{L} (2L+1)\,
\left|1-S_{L}\right|^2\,.
\end{eqnarray}
The cross sections of the  charge exchange
 process $p\bar p\to n\bar n$ are:
\begin{eqnarray}{\label{cex}}
&&\Sigma_{1\mu}^{cex}=\frac{\pi}{p^2}\sum_{L,\,J}\left|{\tilde
R}^{J}_{L\mu}\right|^2\,
,\nonumber\\
&&\Sigma_{00}^{cex}=\frac{\pi}{p^2}\sum_{L} (2L+1)\, \left|{\tilde
S}_{L}\right|^2\,.
\end{eqnarray}

\section{Numerical results}
Using the results obtained above we can discuss the kinetics of the
polarization buildup. Let $\bm P_T$ be the target polarization
vector and $\bm\zeta_T=\bm P_T/P_T$. The arising  polarization $\bm
P_B(t)$ of the antiproton beam is collinear  to $\bm \zeta_T$. The
general solution of the kinetic equation describing the polarization
buildup is given in \cite{MilStr05}. As shown  in this paper, under
certain conditions which are usually fulfilled in
 storage rings, the quantity $P_B(t)$ and the number of
particles in the beam $N(t)$ have the form
\begin{eqnarray}\label{PB}
&&P_B(t)=\tanh\biggl[\frac{t}{2}\,(\Omega_{-}^{out}-\Omega_{+}^{out})\biggr]\,
,\nonumber\\
&&N(t)=\frac{1}{2}N(0)\left[\exp\left(-\Omega_{+}^{out}t\right)+
\exp\left(-\Omega_{-}^{out}t\right)\right]\,,
\end{eqnarray}
where \begin{eqnarray}\label{OmegaPM}
\Omega_{\pm}^{out}&=&nf\left\{\sigma_0\pm P_T\left[
\sigma_1+({\bm\zeta}_T\cdot{\bm
\nu})^2(\sigma_2-\sigma_1)\right]\right\}\, .
\end{eqnarray}
Here $n$ is the  areal density  of the target ,  $f$ is a revolution
frequency of the beam. The function $P_B(t)$ in Eq.(\ref{PB})
contains $\Omega_{\pm}^{out}$ only in the combination
$\Omega_{-}^{out}-\Omega_{+}^{out}$. For $|\sigma_2|>|\sigma_1|$,
this difference is maximal  at ${\bm\zeta}_T
\parallel{\bm \nu}$. For $|\sigma_2|<|\sigma_1|$, the difference
is maximal at ${\bm\zeta}_T \perp{\bm \nu}$.

Our estimation shows that   $|\Omega_{-}^{out}-\Omega_{+}^{out}|\ll
(\Omega_{-}^{out}+\Omega_{+}^{out})$, and below this relation is
assumed to be fulfilled. Then the beam lifetime, $\tau_b$, due to
the interaction with  a target is
\begin{equation}\label{tb}
\tau_b=2/(\Omega_{-}^{out}+\Omega_{+}^{out})\, .
\end{equation}
 Note that the Figure Of Merit,
$FOM(t)=P_B^2(t)N(t)$, is maximal at $t_0=2\tau_b$
 when the number of
antiprotons is  $N(t_0)\approx 0.14 N(0)$. For the polarization
degree at $t_0$   we have
\begin{eqnarray}\label{PBt0}
P_B(t_0)=\left\{\begin{array}{cl} - 2P_T\, \sigma_1/\sigma_0\,  ,
&\mbox{ if}\,\, {\bm\zeta}_T\cdot{\bm
 \nu}=0\\
 -2P_T\,\sigma_2/\sigma_0\,  ,   &\mbox{ if}\, \,|{\bm\zeta}_T\cdot{\bm
 \nu}|=1
 \end{array}\right.
\end{eqnarray}
The positive (negative) value of $P_B(t_0)$ means that the beam
polarization is parallel (antiparallel) to $\bm\zeta_T$.

\begin{figure}[h]
\includegraphics[scale=0.7]{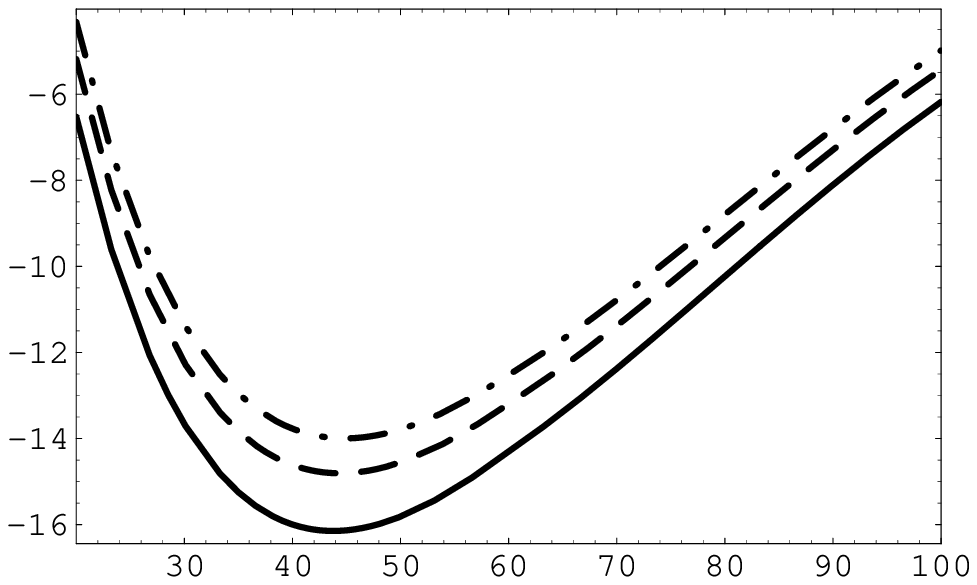}
\hspace{0.5cm}
\includegraphics[scale=0.7]{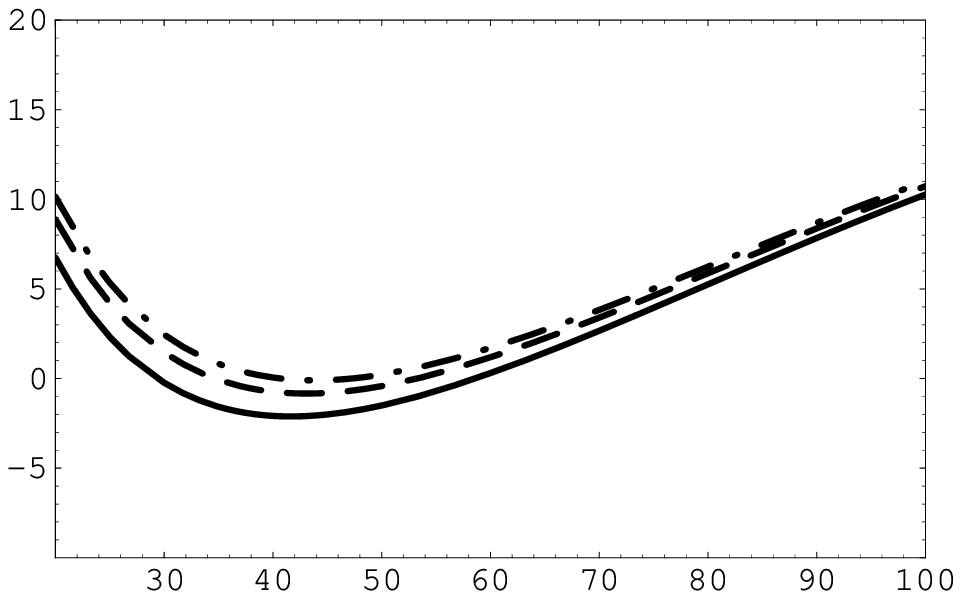}
\begin{picture}(0,0)(0,0)
 \put(-105,-10){ $T$}
  \put(-315,-10){ $T$}
 \put(-215,75){\rotatebox{90}{$\sigma_2$}}
  \put(-430,75){\rotatebox{90}{$\sigma_1$}}
 \end{picture}
\caption{The cross sections $\sigma_1$ (mb) and $\sigma_2$ (mb) as a
function of the kinetic energy $T$ (MeV) in the lab frame. The
acceptance angles in the lab frame are $\theta^{l}_{acc}=10$ mrad
(solid curve), $\theta^{l}_{acc}=20$ mrad (dashed curve), and
$\theta^{l}_{acc}=30$ mrad (dashed-dotted curve) }\label{sig12}
\end{figure}

Predictions for the total  and elastic cross sections, obtained
using the Paris potential with the parameters from
Refs.\cite{paris82,paris94,paris99}, are in good agreement with the
experimental data.  However, the spin-dependent parts, $\sigma_1$
and $\sigma_2$, of the cross section are rather small in comparison
with the total one. Therefore, we can not exclude  that the accuracy
in our predictions for $\sigma_1$ and $\sigma_2$ would be worse
than that for the total cross section. The dependence of $\sigma_1$
and $\sigma_2$ on the antiproton kinetic energy in the lab frame,
$T$, is shown in Fig.\ref{sig12} in the  interval $20\div
100\mbox\,{MeV}$ for a several values of the acceptance angle.

\begin{figure}[h]
\includegraphics[scale=0.68]{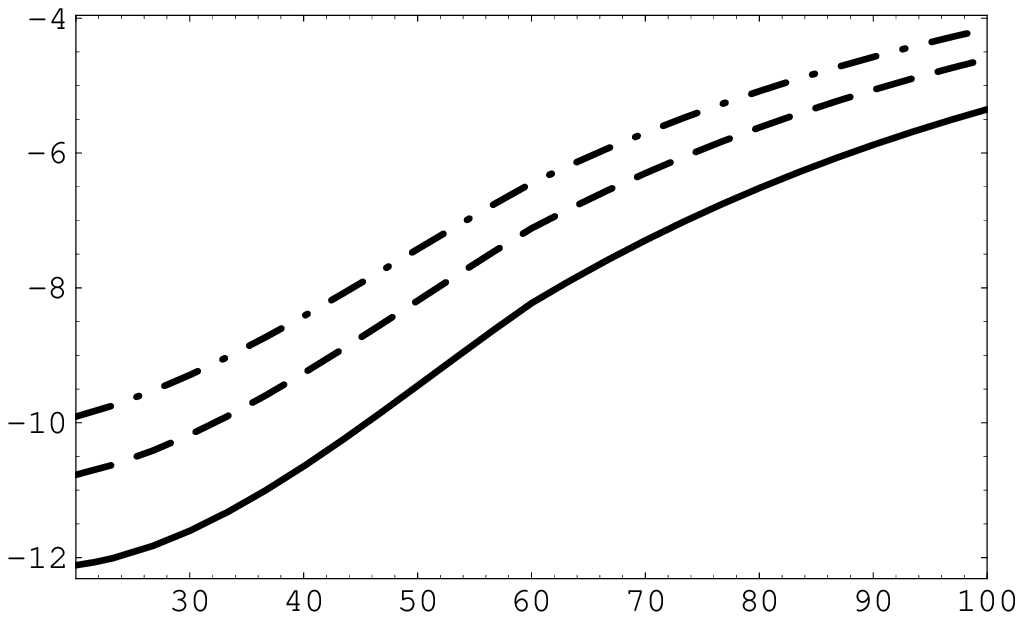}
\hspace{0.5cm}
\includegraphics[scale=0.68]{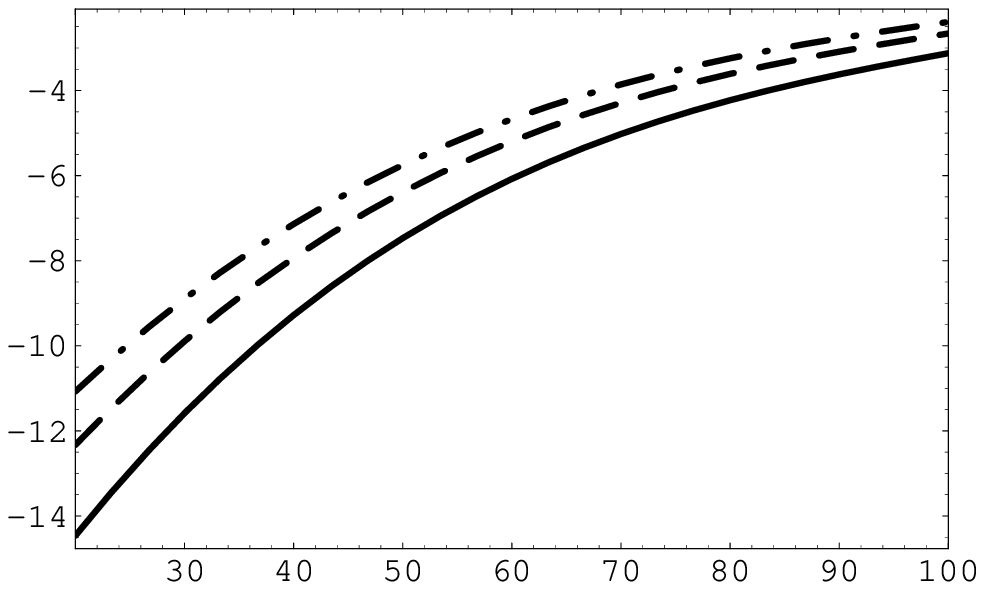}
\begin{picture}(0,0)(0,0)
 \put(-105,-10){ $T$}
  \put(-325,-10){ $T$}
 \put(-222,75){\rotatebox{90}{$\sigma_2^{int}$}}
  \put(-440,75){\rotatebox{90}{$\sigma_1^{int}$}}
 \end{picture}
\caption{The contributions $\sigma_1^{int}$ (mb) and
$\sigma_2^{int}$ (mb) to the cross sections $\sigma_1$ and
$\sigma_2$, respectively, as a function of the kinetic energy $T$
(MeV) in the lab frame.  The acceptance angles  as in
Fig.\ref{sig12}. }\label{sig12int}
\end{figure}

The dependence of $\sigma_{1,2}$ on $\theta_{acc}$ is completely due
to interference of the Coulomb  and the strong elastic amplitudes.
This  interference was very important for describing the proton beam
polarization buildup due to $pp$ scattering, significantly
diminishing both spin-dependent contributions ( $\sigma_{1,2}$) to
the cross section , see Refs.\cite{Meyer94,MilStr05}. Interference
 turns out to be even more important in the $p\bar p$ case, drastically modifying
 $\sigma_{1,2}$ as compared with the pure strong contribution.  The
corresponding quantities, $\sigma_{1,2}^{int}$, are shown in
Fig.\ref{sig12int}. It is seen that $\sigma_1^{int}$ is rather close
to $\sigma_2^{int}$.

\begin{figure}[h]
\includegraphics[scale=0.7]{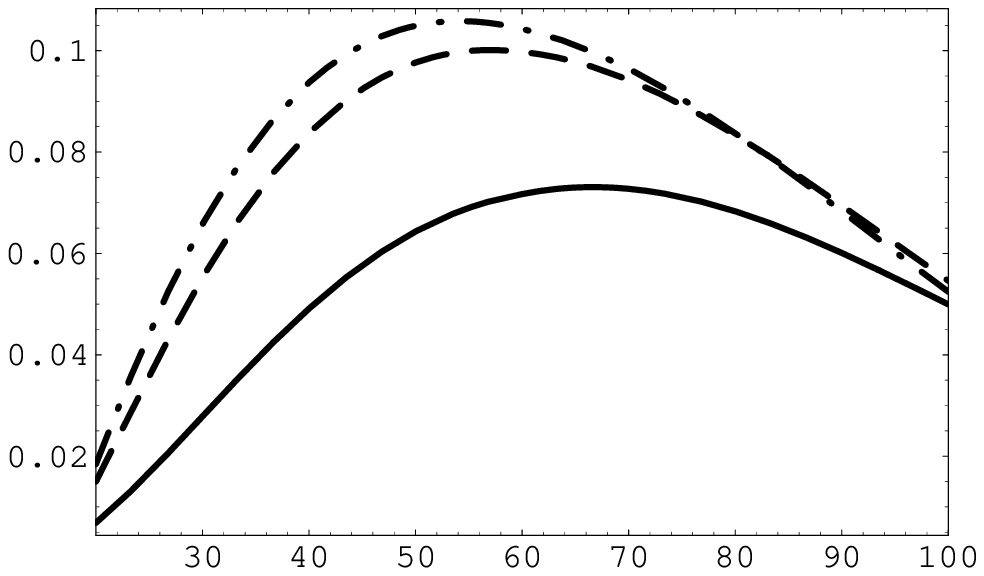}
\hspace{0.5cm}
\includegraphics[scale=0.7]{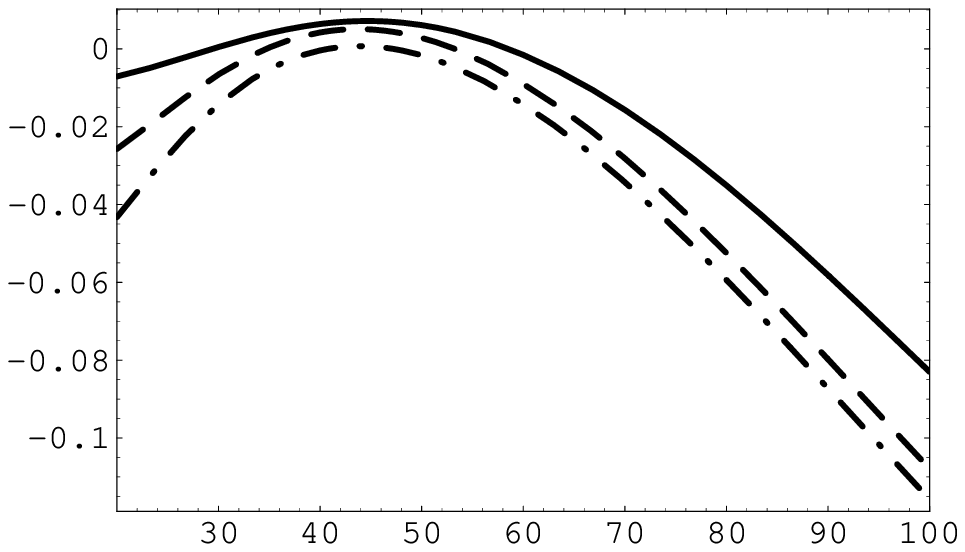}
\begin{picture}(0,0)(0,0)
 \put(-105,-10){ $T$}
  \put(-315,-10){ $T$}
 \put(-220,75){\rotatebox{90}{$P_{\|}$}}
  \put(-435,75){\rotatebox{90}{$P_{\perp}$}}
 \end{picture}
\caption{The polarization  $P_B(t_0)$  at $P_T=1$ as a function of
the kinetic energy $T$ (MeV) in the lab frame for
${\bm\zeta}_T\cdot{\bm \nu}=0$ ($P_{\perp}$) and
$|{\bm\zeta}_T\cdot{\bm \nu}|=1$ ($P_{\|}$) . The acceptance angles
as in Fig.\ref{sig12} .}\label{polar}
\end{figure}

\begin{figure}[h]
\includegraphics[scale=1.1]{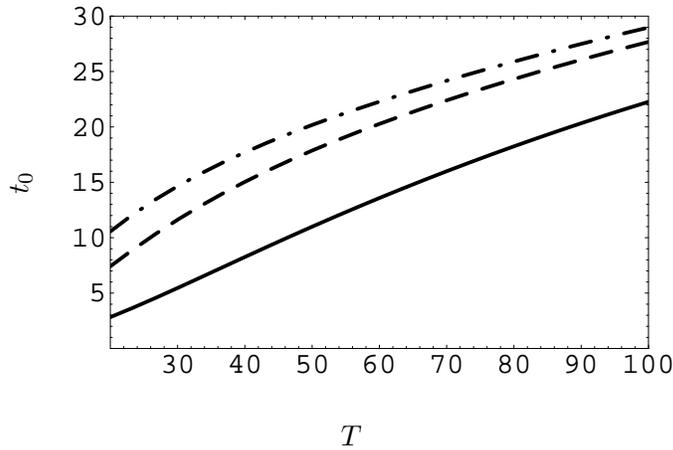}
\begin{picture}(0,0)(0,0)
\put(-135,-10){$T$}
 \put(-260,85){\rotatebox{90}{$t_0$}}
 \end{picture}
\caption{The dependence of  $t_0$ (hour)  on $T$ (MeV) for
$n=10^{14}\,\mbox{cm}^{-2}$, $f=10^6\,\mbox{sec}^{-1}$. The
acceptance angles  as in Fig.\ref{sig12}.}\label{figtau}
\end{figure}

 The dependence of $P_B(t_0)$ on $T$ is shown in
Fig.\ref{polar} for $P_T=1$ , ${\bm\zeta}_T\cdot{\bm \nu}=0$
($P_{\perp}$), and $|{\bm\zeta}_T\cdot{\bm \nu}|=1$ ($P_{\|}$). It
is seen that $P_{\perp}$ has a maximum at energies
$50\div70\,\mbox{MeV}$. The position of this maximum   shifts to the
left with the increasing acceptance angle. The maximal value of
$P_B(t_0)$   also increases with growing $\theta_{acc}$ though this
growth becomes slower at larger $\theta_{acc}$. For
$|{\bm\zeta}_T\cdot{\bm \nu}|=1$, the corresponding  beam
polarization $P_{\|}$ becomes noticeable only for sufficiently large
$T$ where $t_0$ would be too large.
The dependence of  $t_0=2\tau_b$ , Eq.(\ref{tb}), on $T$ is shown in
Fig.\ref{figtau} for $n=10^{14}\,\mbox{cm}^{-2}$,
$f=10^6\,\mbox{sec}^{-1}$, and  several values of the acceptance
angle. It is seen that for these realistic values of the density $n$
and acceptance angle $\theta^{l}_{acc}$, the polarization time  is
rather  reasonable.

 In conclusion, using the Paris nucleon-antinucleon optical
 potential, we have calculated the spin-dependent part of the cross section
 of $p\bar p$ interaction and the corresponding degree of the beam
 polarization. Our results  indicate that a filtering mechanism can
 provide a noticeable   beam  polarization in  a reasonable
 time.

\section*{ACKNOWLEDGMENTS}
 This work was supported in part by  RFBR Grant 05-02-16079.

\end{document}